\documentclass[journal, onecolumn]{IEEEtran}
\usepackage[utf8]{inputenc}

\title{Capacity-achieving codes: a review on double transitivity}
\author{Kirill Ivanov, R\"udiger Urbanke}
\usepackage{amsmath}
\usepackage{cite}
\usepackage{amssymb,amsthm}
\begin{document}
\maketitle
\begin{abstract}Recently it was proved that if a linear code is invariant under the action of a doubly transitive permutation group, it achieves the capacity of erasure channel. Therefore, it is of sufficient interest to classify all codes, invariant under such permutation groups. We take a step in this direction and give a review of all suitable groups and the known results on codes invariant under these groups. It turns out that there are capacity-achieving families of algebraic geometric codes.
\end{abstract}
\section{Introduction}
Consider a permutation group $G$, acting on a set $F$. $G$ is doubly transitive (or 2-transitive) if for any tuple $(i,j,i',i')$ of distinct elements of $F$ there exists a permutation $\pi \in G$, such that $\pi(i)=i,\pi(j)=j'$. All 2-transitive permutation groups are known and classified \cite{dixon1996perm}.

Given a $q$-ary linear code, consider the permutations of its codewords, which keep the code invariant. It can be shown that all such permutations form a group. Kudekar et. al proved that if we can construct a sequence of linear codes with strictly increasing block lengths, converging rates and each of these codes is invariant under a 2-transitive group of permutations, this sequence achieves the capacity of $q$-ary erasure channel under bit-MAP decoding \cite{Kudekar2016RM}. Authors also provide examples of such codes, which are (generalized) Reed-Muller codes, extended narrow-sense BCH codes and quadratic residue codes. However, low-complexity decoding of all these codes is only possible in low- and high-rate regimes. The complete classification of codes, invariant under 2-transitive groups, is an open problem.

From a practical point of view, a theoretically good code should be accompanied with the low-complexity (i.e., polynomial-time) decoding algorithm. In case of erasure channel, bit-MAP and block-MAP decoding can be performed in qubic time via Gaussian elimination, but for general channels the decoding of linear codes is hard \cite{berlekamp1978hard,bruck1990hard}. However, for specific code families (except the Reed-Solomon codes, which are NP-hard to decode \cite{guruswami2005rshard}) the existence of polynomial-time MAP decoding algorithm is unknown. Recently introduced binary polar codes \cite{arikan2008polar}, which derive their structure from Plotkin $(u|u+v)$ construction, are capacity-achieving under quasi-linear successive cancellation (SC) decoding algorithm. SC algorithm can be generalised for decoding of any binary linear code \cite{trifonov2016subcodes}, but for arbitrary codes near-MAP decoding needs complexity that grows exponentially with the code length. 


In this paper, we describe all suitable 2-transitive groups. These groups act on the codeword indices and the codes' sequences should have nontrivial rates, so it is sufficient to study only the nontrivial infinite groups. For each of them, we provide the known results on invariant codes. The most notable discovery is the existence of families of capacity-achieving algebraic geometric codes.

\section{Doubly transitive permutation groups}
There exist only 6 nontrivial infinite families of 2-transitive permutation groups \cite{dixon1996perm}. Below we give the description of each group and its action.
\subsection{Affine groups}
For finite field $GF(q)$, general linear group $GL_d(q)$ is defined as a group of all invertible $d\times d$ matrices over $GF(q)$ and general semi-linear group $\Gamma L_d(q)$ consists of elements of $GL_d(q)$ composed by an automorphism of $GF(q)$.

An affine group action, which is 2-transitive on $q^d$ points, can be represented as $$x\to ax+b,\ x,b\in \mathbb F_q^d,\ a\in G_0\subseteq \Gamma L_d(q),$$ where $G_0$ fixes zero and acts transitively on the set of nonzero elements of $\mathbb F_q^d$. $G_0$ should contain one of three following groups as a subgroup:
\begin{enumerate}
    \item Special linear group $SL_d(q)\subseteq GL_d(q)$, which consists of matrices with determinant 1
    \item Symplectic group $Sp_d(q)\subseteq GL_d(q)$, which is 2-transitive itself when $q=2$ and will be introduced later
    \item Exceptional group of Lie type $G_2(2^m)$
\end{enumerate}

The most famous and studied group is the general affine group $AGL_d(q)$, which corresponds to $G_0=GL_d(q)$.

\subsection{Projective}
Projective (general) linear group $PGL_d(q)$ is a group of $d\times d$ matrices over $GF(q)$ modulo the scalar matrices $B=\text{diag}(a,\dots,a), a\in GF(q)^*$. Its action is 2-transitive on the set of 1-dimensional subspaces of $GF(q)$. Projective special linear group is a 2-transitive subgroup of $PSL_d(q)$, which consists of matrices with determinant 1. For $d=2$, we can equivalently consider the action of projective groups on $GF(q)\cup \{\infty\}$ as 
\begin{align}
    PGL_2(q):&\ x\to \frac{ax+b}{cy+d},\ a,b,c,d \in GF(q)\\
    PSL_2(q):&\ x\to \frac{ax+b}{cy+d},\ a,b,c,d \in GF(q), ad-bc=1
\end{align}
    
\subsection{Symplectic} 
For $m\ge 1$, define matrices $$E=\begin{pmatrix}\mathbf 0_m & \mathbf 1_m\\ \mathbf 0_m & \mathbf 0_m\end{pmatrix},\ F=\begin{pmatrix}\mathbf 0_m & \mathbf 1_m\\ -\mathbf 1_m & \mathbf 0_m\end{pmatrix}=E-E^T.$$ The symplectic group $Sp_{2m}(q)\subset GL_{2m}(q)$ contains all matrices $A$ over $\mathbb F_q^{2m\times 2m}$, such that $A^TFA=F$. 
For $q=2$ its action is 2-transitive on the partitions of set of the following bilinear forms: $$\{\theta_a(u)=uEu^T+uFa^T\},$$ which correspond to $aEa^T=\pm 1$ and have size $2^{2m-1}\pm2^{m-1}$.


\subsection{Unitary, Suzuki and Ree groups}
Projective unitary group $PGU_3(q^2)$ (and its special subgroup) is 2-transitive on $q^3+1$ points, Suzuki group $Sz(q)$ is defined for $q=2^{2m+1}$ and is 2-transitive on $q^2+1$ points and Ree group is defined for $q=3^{2m+1}$ and is 2-transitive on $q^3+1$ points. Actions of these groups are quite cumbersome to describe, but the interested reader can check with the references \cite{dixon1996perm,suzuki1960group,ree1960group} descriptions.

\section{Codes, invariant under doubly transitive groups}
\subsection{Affine codes}
The constructions of codes in case of $G_0$ containing $Sp_d(q),q>2$, $G_2(2^m)$ were not found in the literature. The case $G_0=GL_d(q)$ was studied by Delsarte, who proved the following result \cite{delsarte1970inv}: 
\begin{itemize}
    \item Any nontrivial $q$-ary linear code of length $(q^{rm}-1)/b$ (for some $b$ that divides $q^{rm}-1)$) invariant under $GL_m(q^r)$ is equivalent to a cyclic code;
    \item Any nontrivial $q$-ary linear code of length $q^{rm}$ invariant under $AGL_m(q^r)$ is equivalent to an extended cyclic code;
    \item Any nontrivial $q$-ary linear code of length $q^m$  invariant nder $AGL_m(q)$ is equivalent to an extended Generalized Reed-Muller (GRM) code.
\end{itemize}
\noindent \underline{Remark}: It seems that Delsarte's proof also works if we replace $GL_d(q)$ with $SL_d(q)$.

One of the oldest and the most studied affine-invariant codes are extended BCH and Reed-Muller codes, which are both mentioned by Kudekar et. al in their seminal paper \cite{Kudekar2016RM}. The examples of other affine-invariant codes are given e.g. in \cite{guo2013affine}, although authors mostly concentrate on the rates close to 1.

\subsection{Projective codes}
The most famous family of codes, invariant under $PSL_2(q)$, are $l$-ary extended quadratic residue (QR) codes, which are defined for prime $q$ and $l$ being a quadratic residue modulo $q$. In fact, QR codes are the only binary and quaternary linear codes invariant under $PSL_2(q)$ \cite{ding2017inv}. Generalization of QR codes for $q$ being a prime power, which are defined for $(q,l)=1$ was done by van Lint \cite{vanlint1979gqr}. Another example of ternary codes is given by Pless \cite{pless1972pgl} for prime power $q\equiv -1 \mod 3$. All these codes achieve the capacity of $BEC(1/2)$. 

A remarkable result is given by Korchmáros and Speziali, who used the Hermitian curves to construct codes of length $q^3-q$ for odd prime power $q$, invariant under $PGL_2(q)$ \cite{korchmaros2017hermpgl}. It can be verified that for any $r\in (0,1)$ there exists a sequence of such codes with rates converging to $r$.

Ding et al. proved that for $q=2^m$, there are no nontrivial $2^h$-ary codes, invariant under $PGL_2(q)$ \cite{ding2020projective}. The only result for $d>2$ is given by Malevich and Willems, who verified by exhaustive search that there are no nontrivial binary codes invariant under the action of $PSL_d(q)$ for $(d,q)=(4,3)$, $(4,7)$, $(4,11)$, $(8,3)$ \cite{mal2014aut}.


\subsection{Symplectic codes}
An action of the symplectic group is mostly studied in the framework of combinatorial designs. $t$-design is a set of points and a set of blocks, such that each point appears in the same number of blocks and each $t$-element subset of points appears in the same number of blocks. A linear code from design is spanned by the rows of its incidence matrix (or alternatively its dual). The main result here is given by McGuire \cite{mcguire1997quasi}, who constructed some binary $(2^{2m-1}\pm 2^{m-1}, 2m+1)$ codes from 2-designs. However, it is easy to see that their rates are converging to zero. 

\subsection{Unitary, Suzuki and Ree codes}
The actions of unitary, Suzuki and Ree groups have long-studied connection with algebraic geometric (AG) codes. A $q$-ary AG code is typically constructed by choosing a suitable curve over $\mathbb F_q$ and working with the set of its rational points. The considered groups are the automorphism groups of Hermitian, Suzuki and Ree curves, respectively. The descriptions of these curves can be found, e.g., in \cite{montanucci2017curves}. 

Korchmáros and Nagy constructed $PGU_3$-invariant $q^6$-ary codes of length $q^9-q^3$ \cite{korchmaros2020hermitian}, but the invariance is proved only for rates converging to zero. Eid et al. construct $q^4$-ary Suzuki-invariant codes of length $q^4+\sqrt 2 q^{2.5}(q-1)-q^2$ \cite{eid2014suzuki} and dimension $l(q^2+1)-q_0(q-1)+1$ for $l\le q^2-1$. Extensions of Suzuki and Ree curves are studied in \cite{montanucci2017suzukiree}, where authors introduce $q^4$-ary Suzuki-invariant codes of length $q^5-q^4+q^3-q^2$ and dimension $l(q^2+1)-(q^3-2q^2+q-2)/2$ for $l\le q^3-q^2$ and $q^6$-ary Ree-invariant codes of length $q^7-q^6+q^4-q^3$ and dimension $l(q^3+1)-(q^4-2q^3+q-2)/2$ for $l\le q^4-q^3$. It can be verified that for any $r\in (0,1)$ there exist sequences of such codes with rates converging to $r$.

Malevich and Willems checked by exhaustive search that there are no nontrivial binary codes invariant under the action of $PGU_3(7)$ \cite{mal2014aut} and Brooke enumerated all codes, invariant under $PSU_3(3)$ \cite{brooke1985psu3}.

\section{Conclusion}
In this review, we investigate the known code families, which are invariant under various doubly transitive permutation groups and as a consequence achieve capacity of erasure channels under bit-MAP decoding. Besides the Reed-Muller, eBCH and QR codes, mentioned in the seminal paper, there are not many other suitable code sequences. Surprisingly, we found several constructions of algebraic geometric codes, that satisfy the requirements on rates and automorphism groups and therefore are capacity-achieving.

\bibliographystyle{IEEEtran}
\bibliography{references}

\end{document}